\documentclass[12pt]{article}
\usepackage{amssymb,amsmath,epsfig}
\allowdisplaybreaks

\begin{document}

\title{\bf Charged Anisotropic Models with Complexity-free Condition}
\author{M. Sharif$^1$ \thanks{msharif.math@pu.edu.pk} and Tayyab Naseer$^{1,2}$ \thanks{tayyabnaseer48@yahoo.com}\\
$^1$ Department of Mathematics and Statistics, The University of Lahore,\\
1-KM Defence Road Lahore, Pakistan.\\
$^2$ Department of Mathematics, University of the Punjab,\\
Quaid-i-Azam Campus, Lahore-54590, Pakistan.}

\date{}
\maketitle

\begin{abstract}
This paper uses the definition of complexity for a static
spherically symmetric spacetime and extends it to the case of
charged distribution. We formulate the Einstein-Maxwell field
equations corresponding to the anisotropic interior and calculate
two different mass functions. We then take Reissner-Nordstr\"{o}m
metric as an exterior spacetime to find the matching conditions at
the spherical boundary. Some scalars are developed from the
orthogonal splitting of the curvature tensor, and we call one of
them, i.e., $\mathcal{Y}_{TF}$ as the complexity factor for the
considered setup. Further, the three independent field equations are
not enough to close the system, therefore, we adopt the
complexity-free condition. Along with this condition, we consider
three constraints that lead to different models. We also present the
graphical interpretation of the resulting solutions by choosing some
particular values of parameters. We conclude that the models
corresponding to $p_r=0$ and a polytropic equation of state show
viable and stable behavior everywhere.
\end{abstract}

\section{Introduction}

General theory of relativity ($\mathbb{GR}$) is regarded as the most
acceptable theory in the scientific community among all theories to
describe the gravitational field. The corresponding field equations
establish a relationship between the geometry of spacetime structure
and matter distribution, explained by the Einstein tensor and the
energy-momentum tensor ($\mathbb{EMT}$), respectively. This set of
differential equations can be solved either analytically or
numerically (initial and boundary conditions are needed in this
case) according to the situation under consideration. Despite this,
it is necessary to provide some extra information associated with
the local physics to solve such a system. According to some recent
developments \cite{1}-\cite{3}, several physical phenomena have been
observed while studying compact structures that may cause to deviate
the isotropy and local pressure anisotropy inside those systems.
Furthermore, there exist multiple factors such as inhomogeneous
density, shear and dissipation, etc., which make the isotropic
pressure condition unstable \cite{4}.

In this context, the field equations provide a set of three
independent differential equations for anisotropic sphere, engaging
five unknowns (metric potentials and matter functions). Therefore,
it is essential to provide two extra conditions in terms of an
experiential assumption comprising any physical parameter, or an
equation of state to find their solution \cite{5,6}. In particular,
various authors considered a polytropic equation of state \cite{7,8}
to explore physical attributes of a white dwarf \cite{9}. This study
has been generalized for the case of anisotropic spherical systems
in the framework of $\mathbb{GR}$ \cite{10}-\cite{14}. On the other
hand, the metric potentials have been restricted by applying some
constraints on them such as the Karmarkar condition in which one of
the functions is arbitrarily chosen to generate the total solution
\cite{15}-\cite{20}. An alternate condition to this is taken as the
conformally flat spacetime, leading to the disappearing Weyl tensor
\cite{21}. The whole analysis manifests that the evolution of the
fluid distribution inside a self-gravitating body can be
characterized by a class of scenarios described by multiple
constraints.

In this regard, a most relevant and widely known concept in the
field of astronomy is the complexity that can be adopted to study a
celestial object. The concept of complexity is related to several
physical quantities such as density, heat flux and pressure, etc.,
that make the interior distribution more complex. Several attempts
have been made in multiple disciplines to define this intuitive
notion in terms of aforementioned parameters. Firstly, an
un-unifying definition of complexity was proposed associated with
entropy and data information \cite{22}-\cite{24}. This idea was
initially applied on two physical systems, namely ideal gas and a
perfect crystal. They were observed to be opposite in their
characteristics, however, treated as the complexity-free systems,
resulting in the failure of this definition.

A comprehensive definition of complexity was recently given by
Herrera \cite{25} in terms of density inhomogeneity and anisotropic
pressure. He obtained some scalars by splitting the curvature tensor
orthogonally \cite{26,27} and observed that the above two parameters
are involved in a factor $\mathcal{Y}_{TF}$, thus named it
complexity factor for a static spherical geometry. The basic idea
was that the complexity vanishes either for an isotropic and
homogeneous configuration, or if the effects of anisotropy and
density inhomogeneity cancel each other. Later, this concept has
been generalized to different scenarios (such as static, dynamical
and axially symmetric geometries) to study their corresponding
different evolutionary patterns \cite{28,29}.

Different evolutionary patterns of celestial structures can
efficiently be studied in the presence of electric charge. The
matter source influenced from an electromagnetic field creates a
force in outward direction that counterbalances the inward directed
gravitational pull of a compact star. Consequently, such systems
maintain their stability for a longer time as compared to the
uncharged stars. The solutions to Einstein-Maxwell field equations
and an impact of charge on them have been analyzed
\cite{30}-\cite{35}. Sunzu et al. \cite{36} calculated a mass to
radius relation for a compact strange star and studied how the
presence of charge affects the considered setup. Murad \cite{37}
also explored the effect of an electromagnetic field on the
developed spherically symmetric model. Sharif and his collaborators
\cite{38,39} formulated decoupled anisotropic charged solutions and
found their stable regions.

This paper extends the developed anisotropic spherical models
\cite{40} to the case of charged matter distribution. We organize
this paper as follows. Section \textbf{2} provides $\mathbb{EMT}$
corresponding to anisotropic fluid and the electromagnetic field,
and formulates Einstein-Maxwell field equations for spherical
interior. The junction conditions are also determined by smoothly
matching the interior spacetime with the Reissner-Nordstr\"{o}m
exterior metric at the boundary. The structure scalars are then
presented in section \textbf{3}, from which we choose
$\mathcal{Y}_{TF}$ as the complexity factor for the considered
scenario. Section \textbf{4} displays a brief summary of some
conditions that should be satisfied by a physically realistic model.
Furthermore, we present three different models along with their
graphical interpretation in section \textbf{5}. Lastly, we sum up
our findings in the last section.

\section{Spherical System and Einstein-Maxwell Field Equations}

This section presents the field equations characterizing a static
charged spherically symmetric fluid configuration. In this regard,
we consider a line element representing inner configuration, which
is bounded by a spherical surface $\Sigma$ as
\begin{equation}\label{g6}
ds^2=-e^{\beta_1} dt^2+e^{\beta_2} dr^2+r^2d\theta^2+r^2\sin^2\theta
d\vartheta^2,
\end{equation}
where $\beta_1=\beta_1(r)$ and $\beta_2=\beta_2(r)$. This metric
must satisfy Einstein-Maxwell field equations given in the following
form
\begin{equation}\label{g2}
\mathbb{G}_{\zeta\delta}=8\pi\big(\mathbb{T}_{\zeta\delta}+\mathbb{E}_{\zeta\delta}\big),
\end{equation}
where $\mathbb{G}_{\zeta\delta},~\mathbb{T}_{\zeta\delta}$ and
$\mathbb{E}_{\zeta\delta}$ describe the Einstein tensor,
$\mathbb{EMT}$ for the matter source and the electromagnetic field,
respectively. The $\mathbb{EMT}$ representing anisotropic fluid
configuration has the form
\begin{equation}\label{g5}
\mathbb{T}_{\zeta\delta}=(\rho+p_t)\mathcal{K}_{\zeta}\mathcal{K}_{\delta}+p_t
g_{\zeta\delta}+\left(p_r-p_t\right)\mathcal{W}_\zeta\mathcal{W}_\delta,
\end{equation}
where $\rho,~p_r,~p_t,~\mathcal{W}_{\delta}$ and
$\mathcal{K}_{\delta}$ symbolize the energy density,
radial/tangential pressure, four-vector and four-velocity,
respectively. These quantities take the form for the line element
\eqref{g6} as
\begin{equation}\label{g7}
\mathcal{W}^\delta=(0,e^{\frac{-\beta_2}{2}},0,0), \quad
\mathcal{K}^\delta=(e^{\frac{-\beta_1}{2}},0,0,0),
\end{equation}
which satisfy the relations $\mathcal{W}^\delta
\mathcal{K}_{\delta}=0,~\mathcal{W}^\delta \mathcal{W}_{\delta}=1$
and $\mathcal{K}^\delta \mathcal{K}_{\delta}=-1$.

The electromagnetic field can be characterized by $\mathbb{EMT}$ as
\begin{equation}\label{g1}
\mathbb{E}_{\zeta\delta}=-\frac{1}{4\pi}\left[\frac{1}{4}g_{\zeta\delta}\mathbb{F}^{\mu\sigma}\mathbb{F}_{\mu\sigma}
-\mathbb{F}^{\sigma}_{\zeta}\mathbb{F}_{\delta\sigma}\right].
\end{equation}
Here, $\mathbb{F}_{\mu\sigma}=\psi_{\sigma;\mu}-\psi_{\mu;\sigma}$
is the Maxwell field tensor and
$\psi_{\sigma}=\psi(r)\delta^{0}_{\sigma}$ represents the
four-potential. The Maxwell equations satisfied by these entities
can be expressed in tensorial form as
\begin{equation*}
\mathbb{F}^{\mu\sigma}_{;\sigma}=4\pi \jmath^{\mu}, \quad
\mathbb{F}_{[\mu\sigma;\zeta]}=0,
\end{equation*}
where $\jmath^{\mu}=\varrho \mathcal{K}^{\mu}$, $\jmath^{\mu}$ and
$\varrho$ are the current and charge densities, respectively. These
equations become in the current scenario as
\begin{equation}
\psi''+\frac{1}{2r}\big\{4-r(\beta_1'+\beta_2')\big\}\psi'=4\pi\varrho
e^{\frac{\beta_1}{2}+\beta_2},
\end{equation}
where $'=\frac{\partial}{\partial r}$. By integrating the above
equation, we have
\begin{equation}
\psi'=\frac{s}{r^2}e^{\frac{\beta_1+\beta_2}{2}}.
\end{equation}
Here, $s=\int_0^r \varrho e^{\frac{\beta_2}{2}}\bar{r}^2d\bar{r}$
demonstrates the total charge inside spherical geometry \eqref{g6}.
The non-vanishing components of $\mathbb{EMT}s$ \eqref{g5} and
\eqref{g1} are
\begin{eqnarray}\nonumber
&\mathbb{T}_{00}=\rho e^{\beta_1}, \quad
\mathbb{T}_{11}=p_re^{\beta_2}, \quad
\mathbb{T}_{22}=p_tr^2=\frac{\mathbb{T}_{33}}{sin^2\theta},\\\nonumber
&\mathbb{E}_{00}=\frac{s^2e^{\beta_1}}{8\pi r^4}, \quad
\mathbb{E}_{11}=-\frac{s^2e^{\beta_2}}{8\pi r^4}, \quad
\mathbb{E}_{22}=\frac{s^2}{8\pi
r^2}=\frac{\mathbb{E}_{33}}{sin^2\theta}.
\end{eqnarray}

The Einstein-Maxwell field equations \eqref{g2} for spherical model
\eqref{g6} are calculated as
\begin{align}\label{g8}
8\pi\rho+\frac{s^2}{r^4}&=\frac{1}{r^2}-e^{-\beta_2}\left(\frac{1}{r^2}-\frac{\beta_2'}{r}\right),\\\label{g9}
8\pi{p}_r-\frac{s^2}{r^4}&=e^{-\beta_2}\left(\frac{1}{r^2}+\frac{\beta_1'}{r}\right)-\frac{1}{r^2},\\\label{g10}
8\pi{p}_t+\frac{s^2}{r^4}&=\frac{e^{-\beta_2}}{4}\left[\beta_1'^2-\beta_1'\beta_2'+2\beta_1''-\frac{2\beta_2'}{r}
+\frac{2\beta_1}{r}\right].
\end{align}
The hydrostatic equilibrium condition \big(developed from the
conservation law, i.e.,
$\nabla_\zeta(\mathbb{T}^{\zeta\delta}+\mathbb{E}^{\zeta\delta})=0$\big)
becomes
\begin{align}\label{g12}
&\frac{dp_r}{dr}+\frac{\beta_1'}{2}\big(\rho+p_r\big)-\frac{2}{r}\big(p_t-p_r\big)-\frac{ss'}{4\pi
r^4}=0,
\end{align}
which is also known as the generalized Tolman-Opphenheimer-Volkoff
($\mathcal{TOV}$) equation for charged anisotropic matter
distribution. The mass function in this case becomes
\begin{align}\label{g12a}
m(r)=\frac{r}{2}\bigg(1-e^{\beta_2}+\frac{s^2}{r^2}\bigg),
\end{align}
or in terms of the energy density, it takes the form after combining
with Eq.\eqref{g8} as
\begin{align}\nonumber
m(r)&=4\pi\int_0^r\rho\bar{r}^2d\bar{r}+\int_0^r\frac{ss'}{\bar{r}}d\bar{r}\\\label{g12b}
&=4\pi\int_0^r\rho\bar{r}^2d\bar{r}+\frac{1}{2}\int_0^r\frac{s^2}{\bar{r}^2}d\bar{r}+\frac{s^2}{2r}.
\end{align}
Equation \eqref{g9} provides the value of differential of temporal
metric potential in terms of the mass \eqref{g12a} as
\begin{align}\label{g12c}
\beta_1'=\frac{2\big(4\pi
p_rr^4+mr-s^2\big)}{r\big(r^2-2mr+s^2\big)},
\end{align}
which after substitution in Eq.\eqref{g12} yields
\begin{align}\label{g12d}
&\frac{dp_r}{dr}+\frac{\big(4\pi
p_rr^4+mr-s^2\big)}{r\big(r^2-2mr+s^2\big)}\big(\rho+p_r\big)+\frac{2\Pi}{r}-\frac{ss'}{4\pi
r^4}=0,
\end{align}
where the anisotropic factor is defined as $\Pi=p_r-p_t$.

Since the interior distribution is influenced from the
electromagnetic field, so the most suitable geometry representing
the exterior spacetime is the Reissner-Nordstr\"{o}m solution given
by
\begin{equation}\label{g15}
ds^2=-\bigg(1-\frac{2\bar{\mathcal{M}}}{r}+\frac{\bar{\mathcal{S}}^2}{r^2}\bigg)dt^2
+\bigg(1-\frac{2\bar{\mathcal{M}}}{r}+\frac{\bar{\mathcal{S}}^2}{r^2}\bigg)^{-1}dr^2+r^2d\theta^2+r^2\sin^2\theta
d\vartheta^2,
\end{equation}
where $\bar{\mathcal{M}}$ and $\bar{\mathcal{S}}$ are the total mass
and charge of the exterior. The continuity of fundamental forms of
the matching conditions at the boundary ($r=r_\Sigma=\mathcal{R}$)
yields
\begin{align}\label{g16}
e^{\beta_1}&{_=^\Sigma}1-\frac{2\bar{\mathcal{M}}}{\mathcal{R}}+\frac{\bar{\mathcal{S}}^2}{\mathcal{R}^2},\\\label{g17}
e^{-\beta_2}&{_=^\Sigma}1-\frac{2\bar{\mathcal{M}}}{\mathcal{R}}+\frac{\bar{\mathcal{S}}^2}{\mathcal{R}^2},\\\label{g18}
p_r&{_=^\Sigma}0, \quad s{_=^\Sigma}\bar{\mathcal{S}}.
\end{align}
Equations \eqref{g16}-\eqref{g18} play highly significant role such
that the smooth matching between the metrics \eqref{g6} and
\eqref{g15} would not be possible without taking them into the
account.

\section{Structure Scalars and Complexity of Compact Sources}

The concept of the complexity of a self-gravitating system has
become a topic of great interest in the field of astrophysics. Among
several definitions of the complexity in the literature, it was
proposed that a structure having isotropic and homogeneous
distribution must have a zero complexity. In the light of this, one
can think that a complexity factor in fact measures how the
inhomogeneous density and pressure anisotropy are related to each
other. A definition in terms of these two factors was firstly
proposed by Herrera \cite{25} by splitting the Riemann tensor
orthogonally \cite{26,27}, and then chosen one of the resulting
scalars as the complexity factor. This section briefly discusses the
procedure to obtain the complexity factor that associates with
different physical parameters. The Riemann tensor can be split into
its trace and trace-free parts through the following equation as
\begin{equation}\label{g23}
\mathbb{R}^{\zeta\varphi}_{\delta\vartheta}=\mathbb{C}^{\zeta\varphi}_{\delta\vartheta}+16\pi
\big(\mathbb{T}^{[\zeta}_{[\delta}\delta^{\varphi]}_{\vartheta]}
+\mathbb{E}^{[\zeta}_{[\delta}\delta^{\varphi]}_{\vartheta]}\big)+8\pi
\mathbb{T}\left(\frac{1}{3}\delta^{\zeta}_{[\delta}\delta^{\varphi}_{\vartheta]}
-\delta^{[\zeta}_{[\delta}\delta^{\varphi]}_{\vartheta]}\right),
\end{equation}
and the two tensors such as $\mathcal{Y}_{\varphi\vartheta}$ and
$\mathcal{X}_{\varphi\vartheta}$ are defined by
\begin{eqnarray}\label{g24}
\mathcal{Y}_{\zeta\delta}&=&\mathcal{R}_{\zeta\varphi\delta\vartheta}\mathcal{K}^{\varphi}\mathcal{K}^{\vartheta},\\\label{g25}
\mathcal{X}_{\zeta\delta}&=&^{\ast}\mathcal{R}^{\ast}_{\zeta\varphi\delta\vartheta}\mathcal{K}^{\varphi}\mathcal{K}^{\vartheta}
=\frac{1}{2}\eta^{\omega\sigma}_{\zeta\varphi}\mathcal{R}^{\ast}_{\omega\sigma\delta\vartheta}
\mathcal{K}^{\varphi}\mathcal{K}^{\vartheta},
\end{eqnarray}
where $\mathcal{R}^{\ast}_{\zeta\varphi\delta\vartheta}
=\frac{1}{2}\eta_{\omega\sigma\delta\vartheta}\mathcal{R}^{\omega\sigma}_{\zeta\varphi}$
and $\eta^{\omega\sigma}_{\zeta\varphi}$ is the Levi-Civita symbol.
These tensors can alternatively be expressed in terms of trace
$(\mathcal{Y}_{T},~\mathcal{X}_{T})$ and trace-free parts
$(\mathcal{Y}_{TF},~\mathcal{X}_{TF})$ as
\begin{eqnarray}\label{g26}
\mathcal{Y}_{\zeta\delta}&=&\frac{h_{\zeta\delta}\mathcal{Y}_{T}}{3}+\bigg(\mathcal{W}_{\zeta}\mathcal{W}_{\delta}
-\frac{h_{\zeta\delta}}{3}\bigg)\mathcal{Y}_{TF},\\\label{g27}
\mathcal{X}_{\zeta\delta}&=&\frac{h_{\zeta\delta}\mathcal{X}_{T}}{3}+\bigg(\mathcal{W}_{\zeta}\mathcal{W}_{\delta}
-\frac{h_{\zeta\delta}}{3}\bigg)\mathcal{X}_{TF}.
\end{eqnarray}
Here,
$h_{\zeta\delta}=g_{\zeta\delta}+\mathcal{K}_{\zeta}\mathcal{K}_{\delta}$
is the projection tensor. Using Eqs.\eqref{g23}-\eqref{g27} and
after some manipulation, we obtain the following scalars as
\begin{eqnarray}\label{g28}
&&\mathcal{X}_{T}=8\pi\rho+\frac{s^2}{r^4},\\\label{g28a}
&&\mathcal{X}_{TF}=-\mathcal{E}-4\pi\Pi+\frac{s^2}{r^4},\\\label{g28b}
&&\mathcal{Y}_{T}=4\pi\big(\rho+3p_r-2\Pi\big)+\frac{s^2}{r^4},\\\label{g28c}
&&\mathcal{Y}_{TF}=\mathcal{E}-4\pi\Pi+\frac{s^2}{r^4},
\end{eqnarray}
where the electric part of the Weyl tensor is is defined by
\begin{equation}\label{g29}
\mathcal{E}=\frac{e^{-\beta_2}}{4}\left[\beta_1''+\frac{\beta_1'^2-\beta_2'\beta_1'}{2}
-\frac{\beta_1'-\beta_2'}{r}+\frac{2(1-e^{\beta_2})}{r^2}\right].
\end{equation}

These scalar functions are associated with physical attributes of a
self-gravitating structure such as homogeneous/inhomogeneous energy
density and anisotropic pressure. Since our goal is to choose the
complexity factor from the above four scalars, so the following form
of $\mathcal{Y}_{TF}$ \eqref{g28c} indicates that only this factor
involves all the required parameters, given by
\begin{equation}\label{g30}
\mathcal{Y}_{TF}=\frac{4\pi}{r^3}\int_0^r\bar{r}^3\rho'd\bar{r}-\frac{3}{r^3}\int_0^r\frac{ss'}{\bar{r}}d\bar{r}
-8\pi\Pi+\frac{2s^2}{r^4}.
\end{equation}
This factor disappears for the isotropic and homogeneous
configuration in the absence of charge. Further, the Tolman (or
active gravitational) mass \cite{aa} is referred as
\begin{equation}\label{g31}
m_T=4\pi\int_0^r\bar{r}^2e^{(\beta_1+\beta_2)/2}(-\mathbb{T}^0_0-\mathbb{E}^0_0+\mathbb{T}^1_1+\mathbb{E}^1_1
+2\mathbb{T}^2_2+2\mathbb{E}^2_2)d\bar{r},
\end{equation}
which helps to explain the energy of the fluid distribution. Equation \eqref{g31} yields an alternative expression,
after some calculations, as
\begin{align}\nonumber
m_T&=(m_T)_\Sigma\bigg(\frac{r}{\mathcal{R}}\bigg)^3+r^3\int_r^\mathcal{R}\frac{e^{(\beta_1+\beta_2)/2}}{\bar{r}^5}\\\label{g32}
&\bigg\{4\pi\bar{r}\int_0^r\tilde{r}^3\rho'd\tilde{r}-3\bar{r}\int_0^r\frac{ss'}{\tilde{r}}d\tilde{r}
-8\pi\bar{r}^4\Pi+4s^2\bigg\}d\bar{r}.
\end{align}
The Tolman mass \eqref{g32} can also be written in terms of scalar
\eqref{g30} as
\begin{align}\label{g33}
m_T&=(m_T)_\Sigma\bigg(\frac{r}{\mathcal{R}}\bigg)^3+r^3\int_r^\mathcal{R}\frac{e^{(\beta_1+\beta_2)/2}}{\bar{r}^5}
\big\{\mathcal{Y}_{TF}\bar{r}^4+2s^2\big\}d\bar{r}.
\end{align}
Equation \eqref{g33} interprets how the active gravitational mass is
influenced by the inhomogeneous density and pressure anisotropy
through the function $\mathcal{Y}_{TF}$. It is important to point
out here that the complexity-free system is not only characterized
by isotropic and homogeneous distribution. Indeed, Eq.\eqref{g30}
also describes such system ($\mathcal{Y}_{TF}=0$) only if the
following condition holds
\begin{equation}\label{g34}
\Pi=\frac{1}{2r^3}\int_0^r\bar{r}^3\rho'd\bar{r}-\frac{3}{8\pi
r^3}\int_0^r\frac{ss'}{\bar{r}}d\bar{r}+\frac{s^2}{4\pi r^4},
\end{equation}
which implies that there exist a class of solutions satisfying this
condition. Since the above condition provides a non-local equation
of state, thus this would be very helpful in constructing the
solutions to the field equations \eqref{g2}.

\section{Some Physical Conditions for the Acceptance of Realistic Models}

Numerous strategies have been found in the literature to solve the
field equations describing physically acceptable compact objects.
However, if these solutions fail to satisfy acceptability
conditions, they are no more relevant to model real compact systems.
Multiple conditions in this regard have been proposed and complied
by various authors \cite{ab,ac}. Some of them are highlighted in the
following.
\begin{itemize}
\item In the interior of a self-gravitating star, the behavior of
radial and temporal metric functions must be finite,
singularity-free and positive.
\item The matter variables (energy density and pressure) must be
finite and maximum at the core ($r=0$) and positive in the whole
domain. Moreover, their derivatives should be zero at $r=0$ and
negative towards the boundary to show monotonically decreasing
trend.
\item For the charged system, the compactness is given by
\begin{align}\nonumber
\mathcal{R}-\sqrt{\mathcal{R}^2-2\mathcal{\bar{M}}\mathcal{R}+\mathcal{\bar{S}}^2}
-\bigg(\mathcal{M}-\frac{\mathcal{\bar{S}}^2}{2\mathcal{R}}\bigg)\leq
\frac{1}{2}\bigg(\mathcal{\bar{M}}-\frac{\mathcal{\bar{S}}^2}{2\mathcal{R}}\bigg),
\end{align}
where
$\mathcal{\bar{M}}-\frac{\mathcal{\bar{S}}^2}{2\mathcal{R}}\neq0$.
The analogous definition of compactness in this case becomes
\begin{align}\nonumber
\frac{\mathcal{\bar{M}}}{\mathcal{R}} \leq
\frac{8}{9}\bigg\{\frac{1}{1+\sqrt{1-\frac{8\gamma^2}{9}}}\bigg\},
\quad \gamma^2=\frac{\mathcal{\bar{S}}^2}{\mathcal{\bar{M}}^2}.
\end{align}
Equivalently, we have \cite{42ga,42gb}
\begin{align}\label{g49}
\frac{1}{\mathcal{R}}\bigg(2\mathcal{\bar{M}}-\frac{\mathcal{\bar{S}}^2}{\mathcal{R}}\bigg)
\leq \frac{8}{9}.
\end{align}
\item The presence of ordinary matter in the interior of a celestial body can be
guaranteed by satisfying some constraints imposed on $\mathbb{EMT}$,
named as the energy conditions. They have the following form in the
current setup
\begin{eqnarray}\nonumber
&&\rho \geq 0, \quad \rho+p_{r} \geq 0,\\\nonumber &&\rho+p_{t} \geq
0, \quad \rho-p_{r} \geq 0,\\\label{g50} &&\rho-p_{t} \geq 0, \quad
\rho+p_{r}+2p_{t} \geq 0.
\end{eqnarray}
The most essential bounds among all the above conditions to be
fulfilled are the dominant energy conditions (i.e., $\rho-p_{r}$ and
$\rho-p_{t}$), which claim $\rho\geq p_{r}$ and $\rho\geq p_{t}$
everywhere.
\item The redshift for the interior configuration is defined as
$Z=e^{-\beta_1/2}-1$. Since this factor depends only on the temporal
metric potential, thus it must decrease towards the boundary. Also,
its value must be less than or equal to $5.211$ at $\Sigma$ to get
an acceptable model \cite{ad}.
\item To check the stability of a celestial system just departed from hydrostatic equilibrium has become a topic
of great interest now a days. Herrera et al. \cite{ae,af} discussed
the notion of cracking which occurs inside the fluid when sign of
the total radial force changes at some particular point. The
cracking can be avoided if the following inequality holds
\begin{align}\label{g51}
-1 \leq v_{st}^{2}-v_{sr}^{2} \leq 0,
\end{align}
where $v_{sr}^{2}=\frac{dp_{r}}{d\rho}$ and
$v_{st}^{2}=\frac{dp_{t}}{d\rho}$ are the radial and tangential
sound speeds, respectively.
\end{itemize}

\section{Models}

There are multiple possible models suggested by different authors to
calculate solutions of the field equations. For example, Herrera
\cite{25} formulated such solutions by considering two constraints
like the Gokhroo and Mehra ansatz and the polytropic equation along
with vanishing complexity factor. Among all the conditions, we adopt
three different models and analyze physical characteristics of the
corresponding developed solutions in the following subsections.

\subsection{Complexity-free Condition with $p_r=0$ Case}

The field equations \eqref{g8}-\eqref{g10} contain six unknowns
($\rho,p_r,p_t,\beta_1,\beta_2,s$), thus we shall require to take
some extra constraints to find their solution. We adopt the known
form of the charge \cite{ag} to reduce one degree of freedom, that
is why we are left with five unknowns to be calculated. For this, we
consider $\mathcal{Y}_{TF}=0$ and $p_r=0$ constraints which
represent a spherical solution characterized only by the tangential
pressure due to Florides \cite{ah}. The later condition provides
from Eq.\eqref{g9} as
\begin{align}\label{g52}
e^{\beta_2}=\frac{r^2\big(1+\beta_1'r\big)}{r^2-s^2}.
\end{align}
Inserting Eqs.\eqref{g8}, \eqref{g10} and \eqref{g52} into
\eqref{g34}, we obtain a second order differential equation for
$\beta_1$ as
\begin{align}\nonumber
&r\beta_1'\big(s^4-2r^3(12\mathcal{Q}+r)-5s^2r^2+\big(r^6-s^2r^4\big)\beta_1''\big)-2r^5(6\mathcal{Q}+r)\beta_1'^2\\\label{g53}
&-12\mathcal{Q}r^3+s^4-5s^2r^2+\big(r^7-s^2r^5\big)\beta_1'^3+\big(2r^6-2s^2r^4\big)\beta_1''=0,
\end{align}
where $\mathcal{Q}=\int_0^r\frac{ss'}{r}dr$.

Equation \eqref{g53} can only be integrated through numerical
technique, whose solution along with \eqref{g52} make both the
metric coefficients known. Figure \textbf{1} exhibits the plots of
metric potentials $e^{\beta_1}$ and $e^{-\beta_2}$ whose behavior is
positive, singularity-free and finite everywhere. Moreover, we
observe that these functions attain expected values at the center,
as $e^{\beta_1(0)}=\mathcal{C}$ (where $\mathcal{C}$ is a positive
constant) and $e^{-\beta_2(0)}=1$. The radius of the surface of a
star is a point where these potentials coincide. In this case, we
find that they coincide at $\mathcal{R}\approx0.12$, so that the
matching conditions give
$$e^{\beta_1(0.12)}=e^{-\beta_2(0.12)}\approx0.87=1-\frac{2\mathcal{\bar{M}}}{\mathcal{R}}
+\frac{\mathcal{\bar{S}}^2}{\mathcal{R}^2},$$ which provides
compactness of the current structures as
\begin{align}\label{g54}
\frac{2\mathcal{\bar{M}}}{\mathcal{R}}
-\frac{\mathcal{\bar{S}}^2}{\mathcal{R}^2}\approx0.13 < \frac{8}{9}.
\end{align}

Figure \textbf{2} (left plot) presents the behavior of energy
density which is maximum at $r=0$ and decreases towards the
hypersurface. Furthermore, the presence of charge makes the system
less dense as compared to the uncharged configuration. Since the
system with vanishing radial pressure is related to Florides
solution, therefore the spherical stability can only be preserved if
the tangential pressure has a positive and increasing behavior
outwards. Figure \textbf{2} (right plot) shows the trend of $p_t$
that is consistent with the required result. The nature of
anisotropy is observed to be zero at the core and opposite from the
tangential pressure towards the boundary (lower plot).

The energy conditions are plotted in Figure \textbf{3} that provide
acceptable behavior, resulting in the viability of this solution.
Figure \textbf{4} (left) manifests the interior redshift that
decreases with increases $r$, and its value at the surface is found
as $Z(0.12)\approx0.96$. This value is much less than its observed
upper limit, i.e., $Z_{\mathcal{R}}=5.211$. The stability is also
checked in Figure \textbf{4} (right) from which we notice that the
developed model, in this case, avoids cracking and retains stability
everywhere.
\begin{figure}\center
\epsfig{file=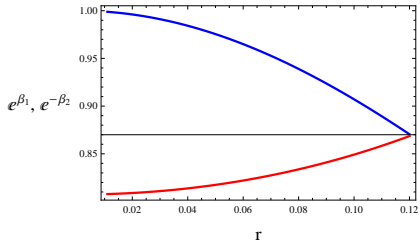,width=0.4\linewidth}
\caption{Plots of $e^{\beta_1}$ (red) and $e^{-\beta_2}$ (blue)
corresponding to Model I.}
\end{figure}
\begin{figure}\center
\epsfig{file=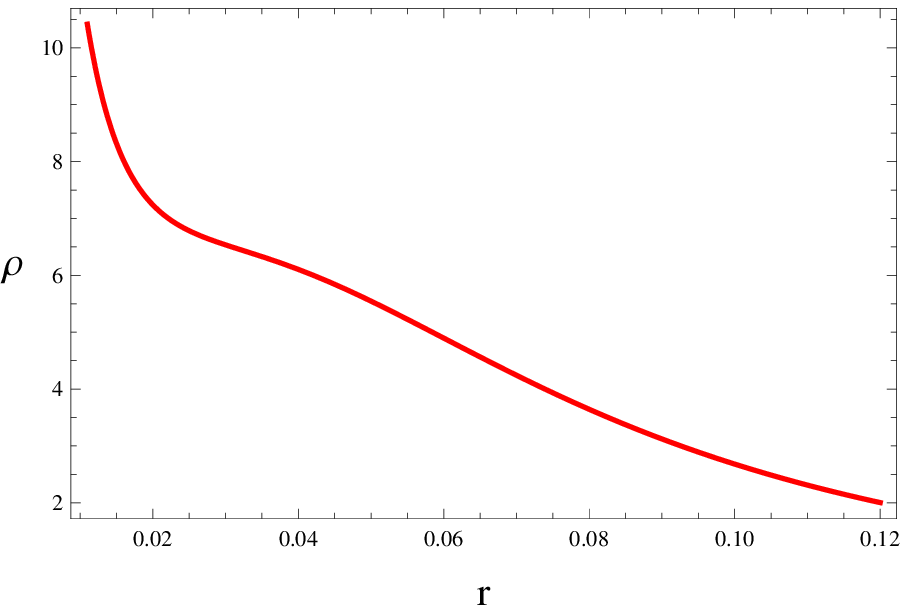,width=0.4\linewidth}\epsfig{file=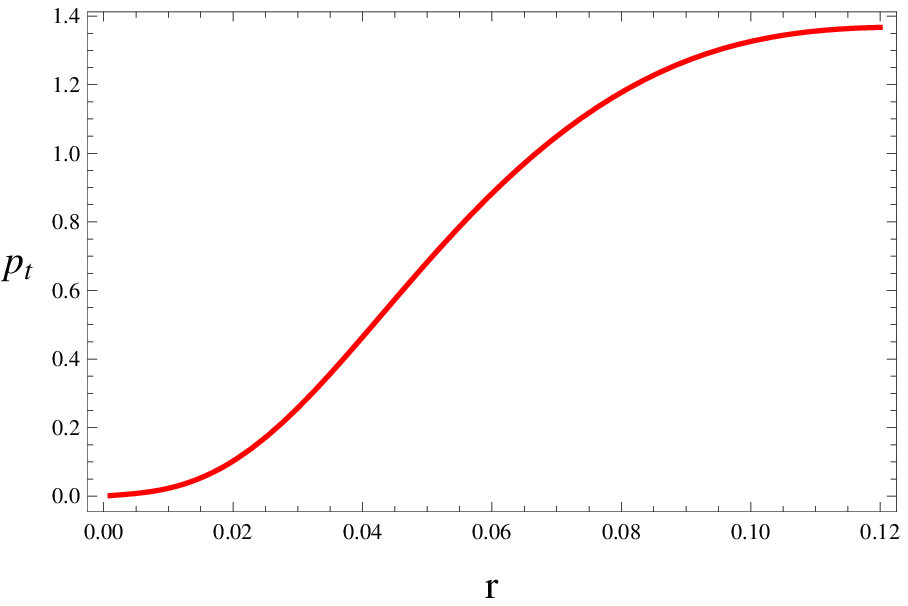,width=0.4\linewidth}
\epsfig{file=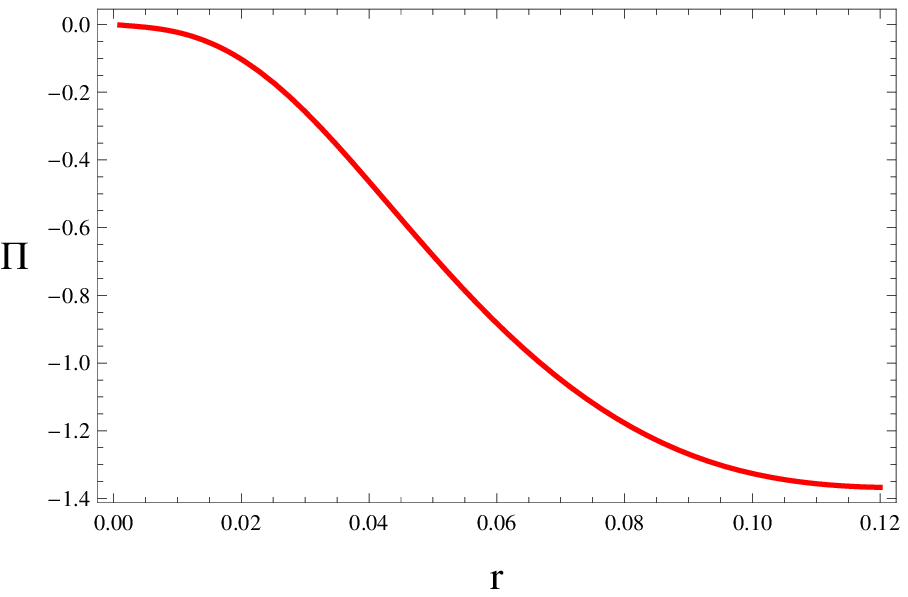,width=0.4\linewidth} \caption{Plots of
energy density, tangential pressure and anisotropy corresponding to
Model I.}
\end{figure}
\begin{figure}\center
\epsfig{file=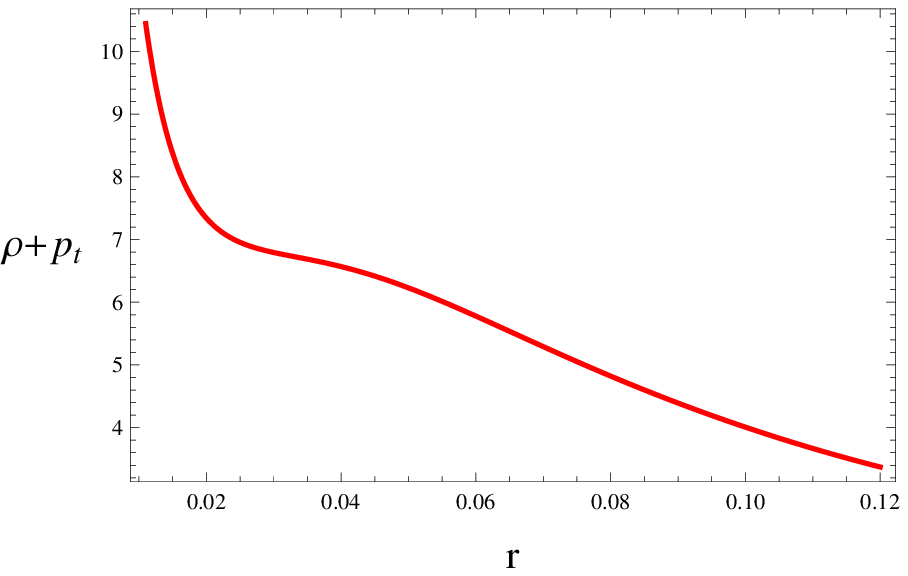,width=0.4\linewidth}\epsfig{file=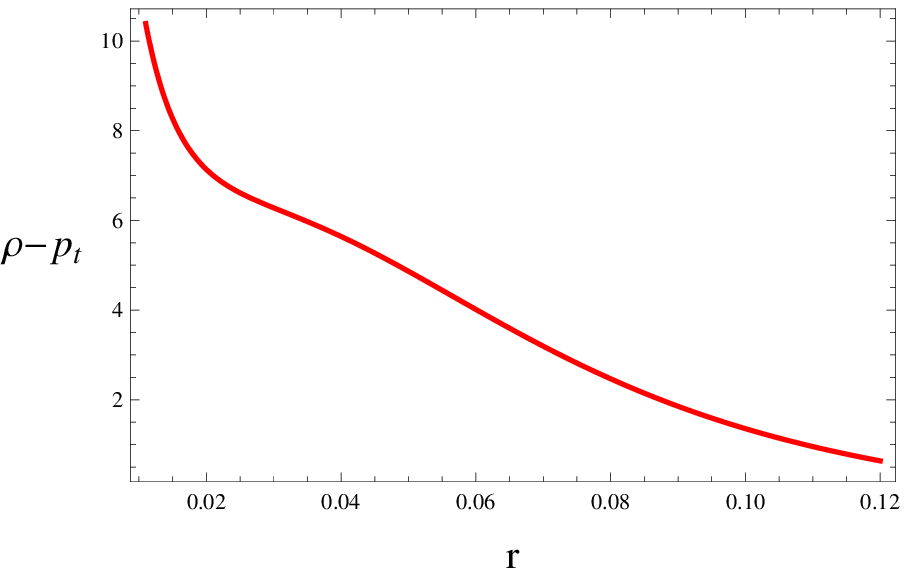,width=0.4\linewidth}
\epsfig{file=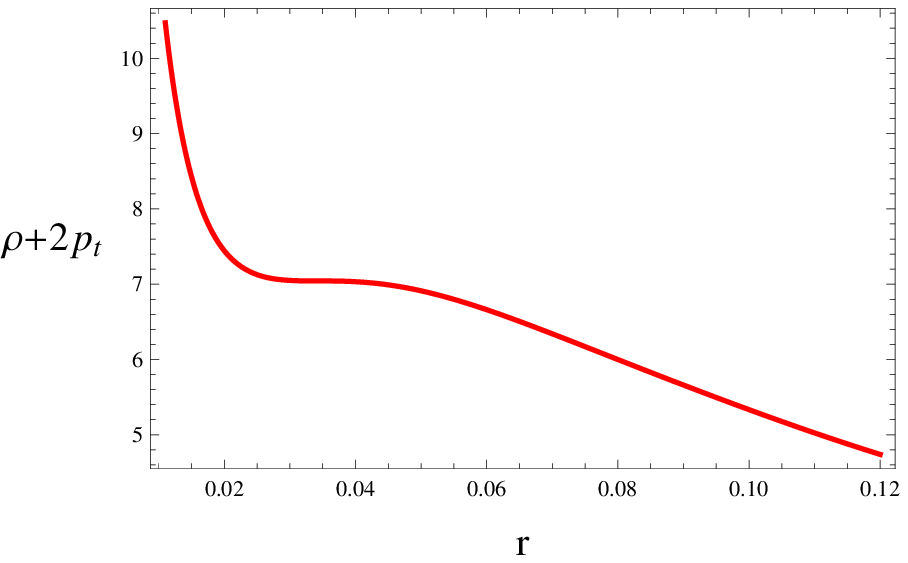,width=0.4\linewidth} \caption{Plots of
energy conditions corresponding to Model I.}
\end{figure}
\begin{figure}\center
\epsfig{file=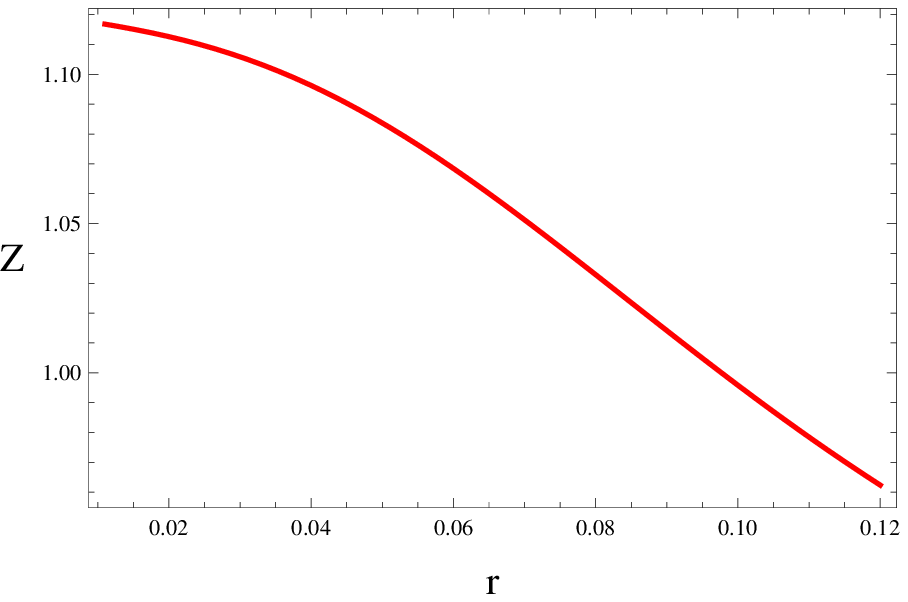,width=0.4\linewidth}\epsfig{file=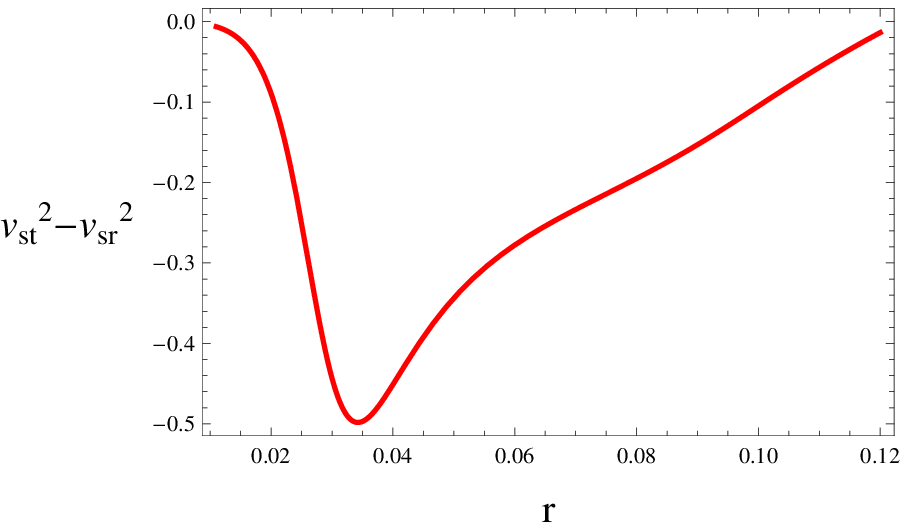,width=0.4\linewidth}
\caption{Plots of redshift and cracking condition corresponding to
Model I.}
\end{figure}

\subsection{Complexity-free Polytrope}

The polytrope corresponding to anisotropic distribution plays
a crucial role in astrophysics and has been discussed thoroughly by
various authors \cite{12}-\cite{14}. Since we need two constraints
to solve the field equations \eqref{g8}-\eqref{g10}, thus a
polytropic equation of state along with complexity-free condition is
taken in this case. A brief discussion was presented on this model
in \cite{25}, however, we analyze the corresponding solution in more
detail through graphical interpretation. We provide the following
two conditions to continue our analysis as
\begin{align}\label{g55}
p_r=\mathcal{H}\rho^\eta=\mathcal{H}\rho^{1+\frac{1}{\mathcal{N}}},
\quad \mathcal{Y}_{TF}=0,
\end{align}
where $\mathcal{H},~\mathcal{N}$ and $\eta$ symbolize the polytropic
constant, polytropic index and polytropic exponent, respectively.

We now introduce some new variables to convert $\mathcal{TOV}$
equation and the mass function in dimensionless form as
\begin{align}\label{g56}
\zeta=\frac{p_{rc}}{\rho_c}, \quad r=\frac{\xi}{\mathcal{B}}, \quad
\mathcal{B}^2=\frac{4\pi\rho_c}{\zeta(\mathcal{N}+1)},\\\label{g57}
\Psi^{\mathcal{N}}=\frac{\rho}{\rho_c}, \quad
\nu(\xi)=\frac{\mathcal{B}^3m(r)}{4\pi\rho_c},
\end{align}
where $c$ in the subscript defines the value of that quantity at the
center. At $r=\mathcal{R}$ (or $\xi(r)=\xi(\mathcal{R})$), we have
$\Psi\big(\xi(\mathcal{R})\big)=0$. Equations \eqref{g12b} and
\eqref{g12d} can be written after substitution of the above
variables as
\begin{align}\nonumber
&\frac{2\Pi\Psi^{-\mathcal{N}}\xi}{p_{rc}\big(\mathcal{N}+1\big)\big(1+\zeta\Psi\big)}\bigg\{1-\frac{2(\mathcal{N}+1)\zeta\nu}{\xi}
+\frac{4\pi\rho_cs^2}{\zeta\xi^2\big(\mathcal{N}+1\big)}\bigg\}+\frac{\xi^2}{1+\zeta\Psi}\\\nonumber
&\times\bigg\{\frac{d\Psi}{d\xi}-\frac{4\pi\rho_cs\Psi^{\mathcal{N}}}{\xi^4\zeta^3\big(\mathcal{N}+1\big)^3}
\frac{ds}{d\xi}\bigg\}\bigg\{1-\frac{2(\mathcal{N}+1)\zeta\nu}{\xi}
+\frac{4\pi\rho_cs^2}{\zeta\xi^2\big(\mathcal{N}+1\big)}\bigg\}\\\label{g58}
&+\zeta\xi^3\Psi^{\mathcal{N}+1}+\nu-\frac{4\pi\rho_cs^2}{\xi\zeta^2\big(\mathcal{N}+1\big)^2}=0,\\\label{g59}
&4\pi\rho_c\frac{d\nu}{d\xi}=4\pi\rho_c\Psi^{\mathcal{N}}\xi^2+\frac{ss'\mathcal{B}^3}{\xi}.
\end{align}
Since there are two first order differential equations \eqref{g58}
and \eqref{g59} in three variables $\Psi,~\nu$ and $\Pi$, thus we
need one more equation to find these unknowns. For this purpose, we
choose $\mathcal{Y}_{TF}=0$ condition which becomes in terms of
variables \eqref{g56} and \eqref{g57} as
\begin{align}\label{g60}
\frac{6\Pi}{\xi}+2\frac{d\Pi}{d\xi}=\rho_c\mathcal{N}\Psi^{\mathcal{N}-1}\frac{d\Psi}{d\xi}+\frac{ss'\mathcal{B}^3}{4\pi\xi^4}
-\frac{s^2\mathcal{B}^4}{2\pi\xi^5}.
\end{align}
\begin{figure}\center
\epsfig{file=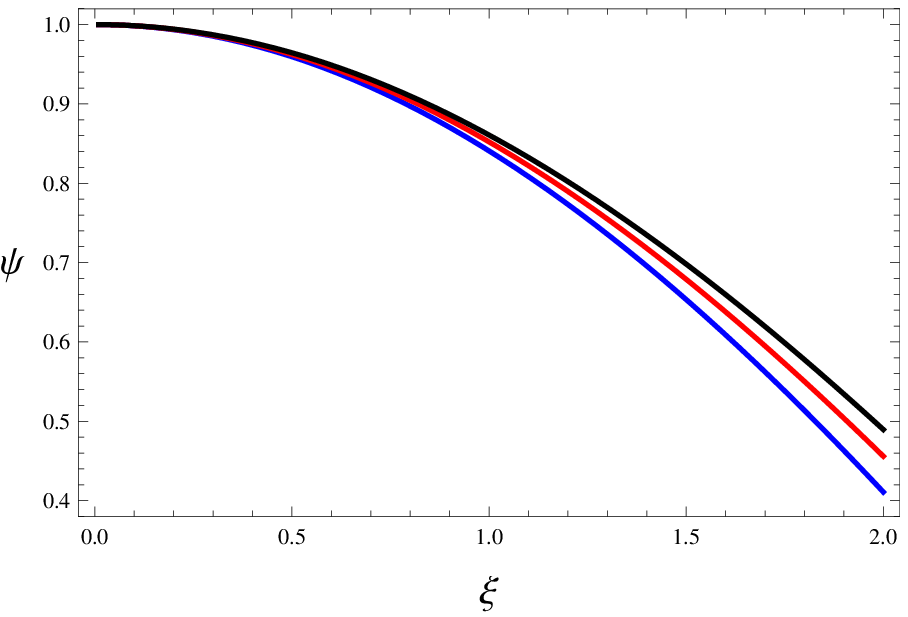,width=0.4\linewidth}\epsfig{file=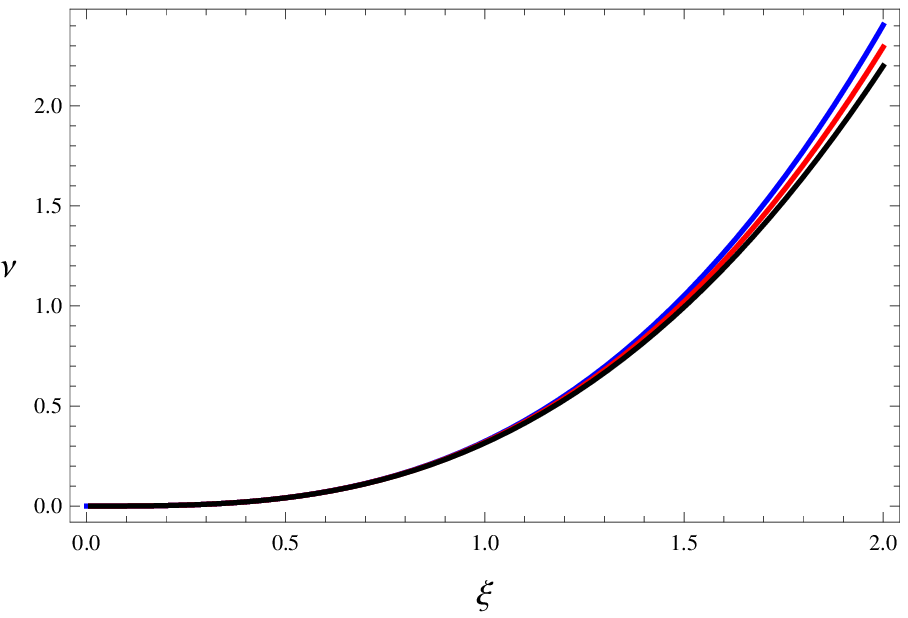,width=0.4\linewidth}
\epsfig{file=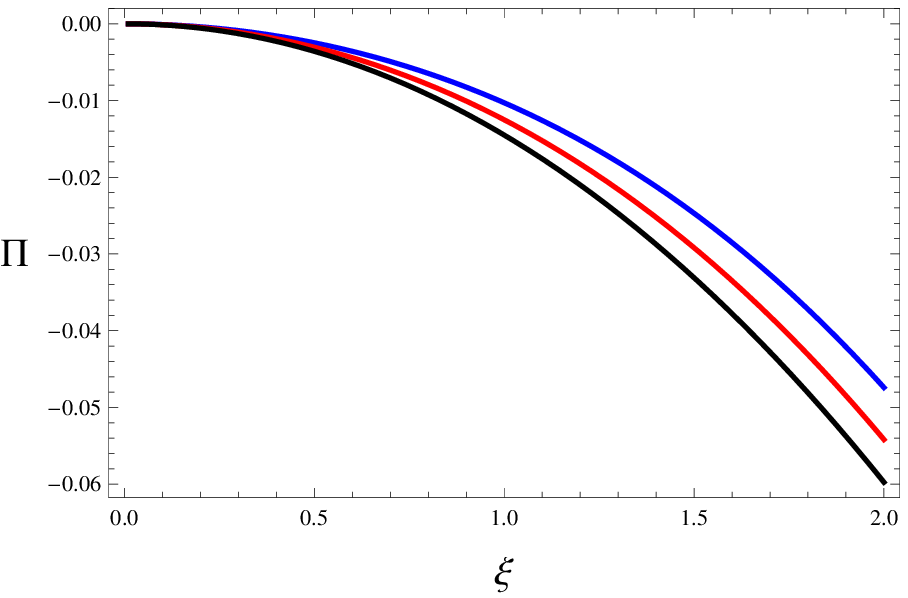,width=0.4\linewidth}
\caption{Plots of $\Psi,~\nu$ and $\Pi$ for $\zeta=0.1$, $\rho_c=1$
and $\mathcal{N}=0.3$ (blue), $\mathcal{N}=0.4$ (red), and
$\mathcal{N}=0.5$ (black) corresponding to Model II.}
\end{figure}

We now have three equations in three unknowns, so a unique solution
can be easily obtained by solving them numerically with the
following conditions
\begin{align}\nonumber
\Pi(0)=0, \quad \nu(0)=0, \quad \Psi(0)=1.
\end{align}
We fix the value of $\zeta$ and observe the graphical behavior of
the above three unknowns as a function of $\xi$ by varying the
parameter $\mathcal{N}$. Figure \textbf{5} (left) shows the behavior
of $\Psi$ (i.e., energy density) which is maximum at $\xi=0$ and
decreases outwards, indicating well-behaved polytropes. The mass of
the considered distribution manifests a monotonically increasing
trend and we observe it in an inverse relation with $\mathcal{N}$
(right plot). The lower plot exhibits anisotropy which disappears at
the core and shows negative behavior towards the surface for all the
chosen values of $\mathcal{N}$.

The geometric sector for the solution corresponding to constraints
\eqref{g55} is plotted in Figure \textbf{6}. We observe the
acceptable behavior of both the metric potentials. Moreover, we get
$e^{\beta_1(0)}=\mathcal{C}$ and $e^{-\beta_2(0)}=1$ at $\xi=0$, and
they coincide at $\xi=2$ which is the surface of a star. We are now
allowed to calculate the compactness factor. For $\mathcal{N}=0.3$,
we have
$$e^{\beta_1(2)}=e^{-\beta_2(2)}\approx0.61=1-\frac{2\mathcal{\bar{M}}}{\mathcal{R}}
+\frac{\mathcal{\bar{S}}^2}{\mathcal{R}^2},$$ which delivers the
compactness as
\begin{align}\label{g61}
\frac{2\mathcal{\bar{M}}}{\mathcal{R}}
-\frac{\mathcal{\bar{S}}^2}{\mathcal{R}^2}\approx0.39 < \frac{8}{9}.
\end{align}
Figure \textbf{7} shows the plots of matter variables such as
density and pressure with respect to a new dimensionless coordinate
$\xi$ for certain values of parameters in Eqs.\eqref{g56} and
\eqref{g57}. We observe their maximum values at the center and
minimum at the boundary following decreasing trend. The radial
pressure becomes zero at $\xi=2$ (upper left plot). Although the
density is observed to be much greater than radial/tangential
pressures, all the energy conditions shown in Figure \textbf{8}
characterize a viable model.

The interior redshift for this model is displayed in Figure
\textbf{9} (left), which shows a decreasing trend with the increase
in $\xi$. Also at $\xi=2$, its value becomes $Z(2)\approx1.48 <
Z_{\mathcal{R}}$, as expected. We also analyze the stable region in
the right plot and find that the cracking condition takes negative
values everywhere, pointing out the stability of the developed
solution unlike the uncharged distribution \cite{40}.
\begin{figure}\center
\epsfig{file=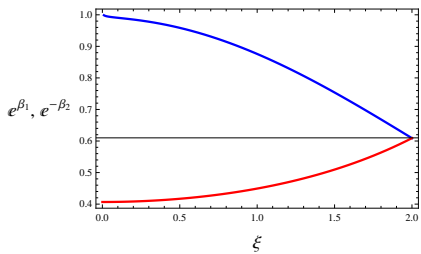,width=0.4\linewidth}
\caption{Plots of $e^{\beta_1}$ (red) and $e^{-\beta_2}$ (blue) for
$\zeta=0.1$, $\rho_c=1$, $\mathcal{H}=0.9$ and $\mathcal{N}=0.3$
corresponding to Model II.}
\end{figure}
\begin{figure}\center
\epsfig{file=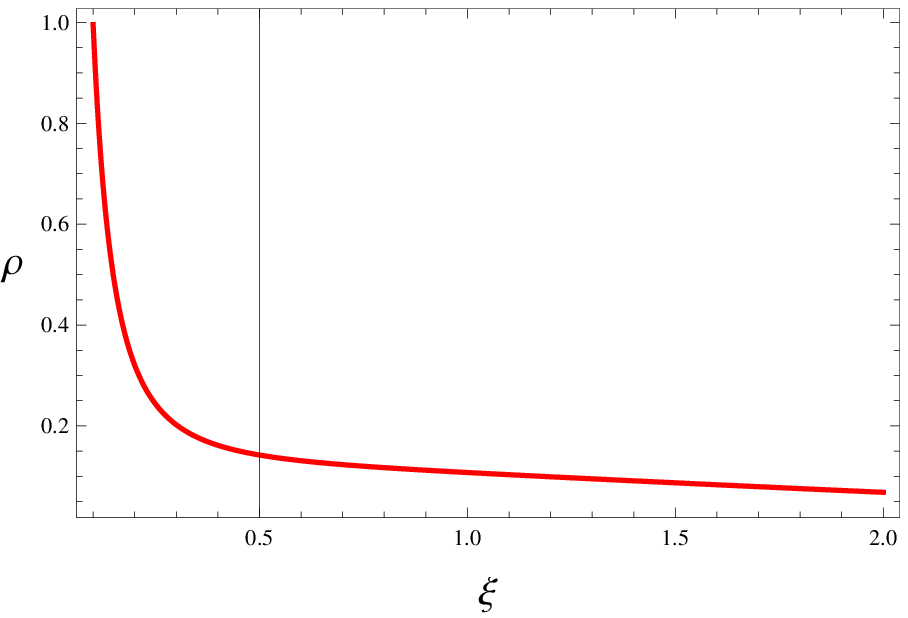,width=0.4\linewidth}\epsfig{file=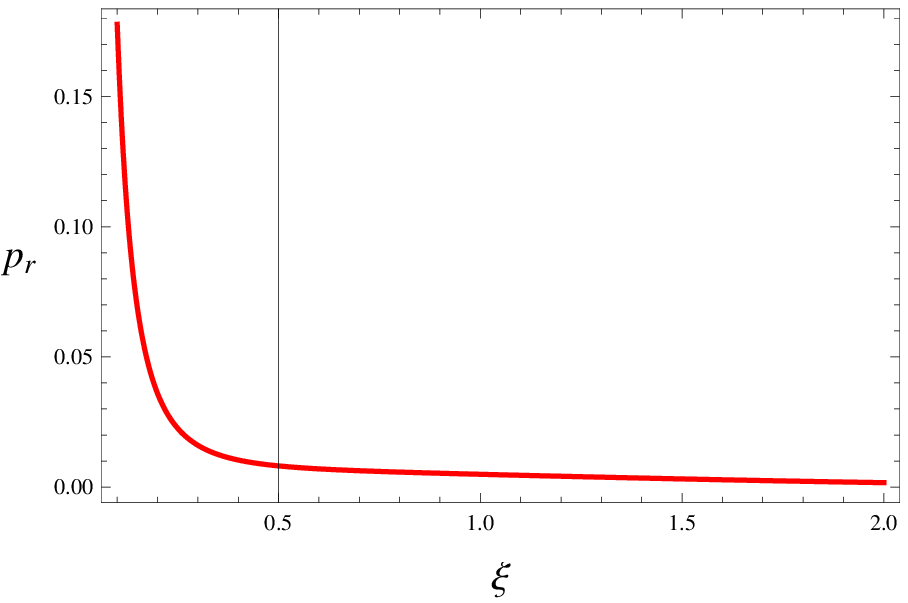,width=0.4\linewidth}
\epsfig{file=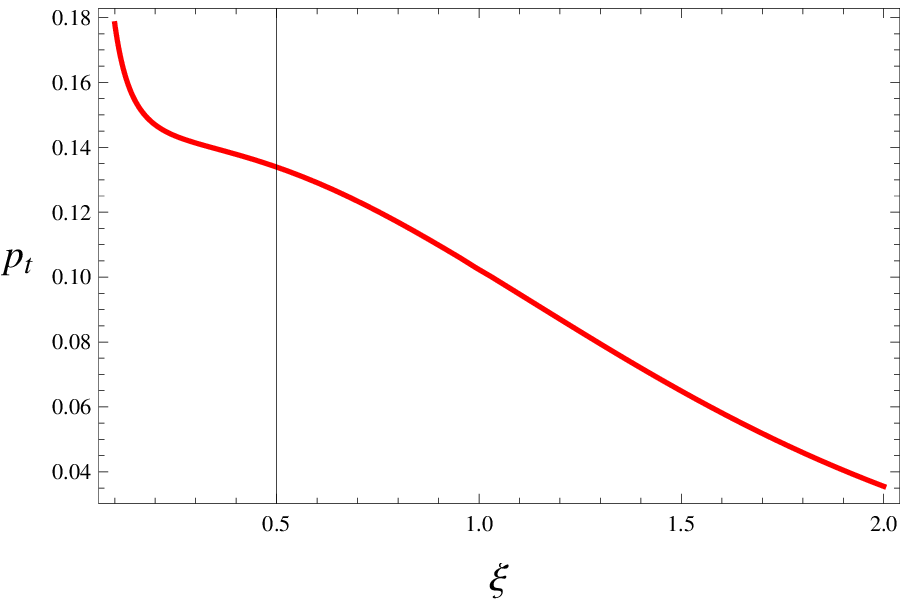,width=0.4\linewidth} \caption{Plots of energy
density, radial and tangential pressures for $\zeta=0.1$,
$\rho_c=1$, $\mathcal{H}=0.9$ and $\mathcal{N}=0.3$ corresponding to
Model II.}
\end{figure}
\begin{figure}\center
\epsfig{file=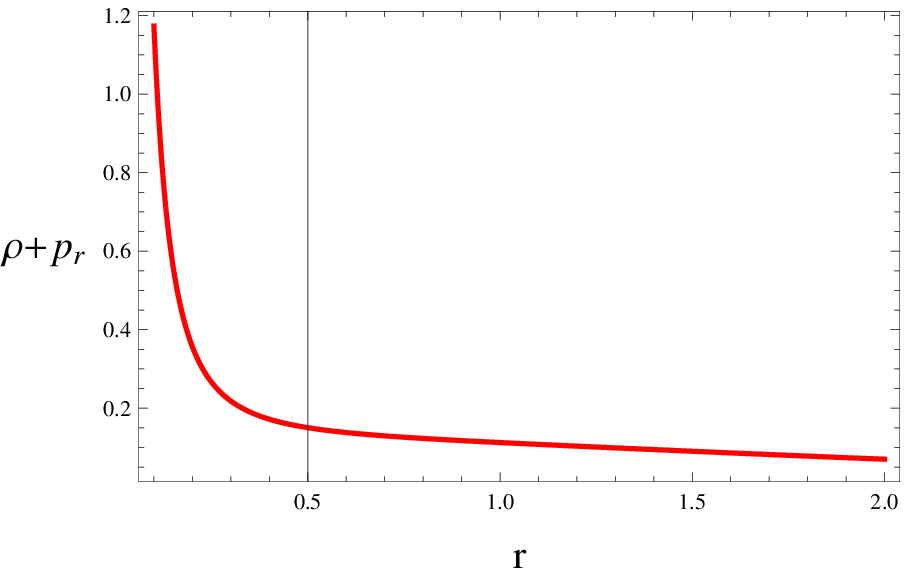,width=0.4\linewidth}\epsfig{file=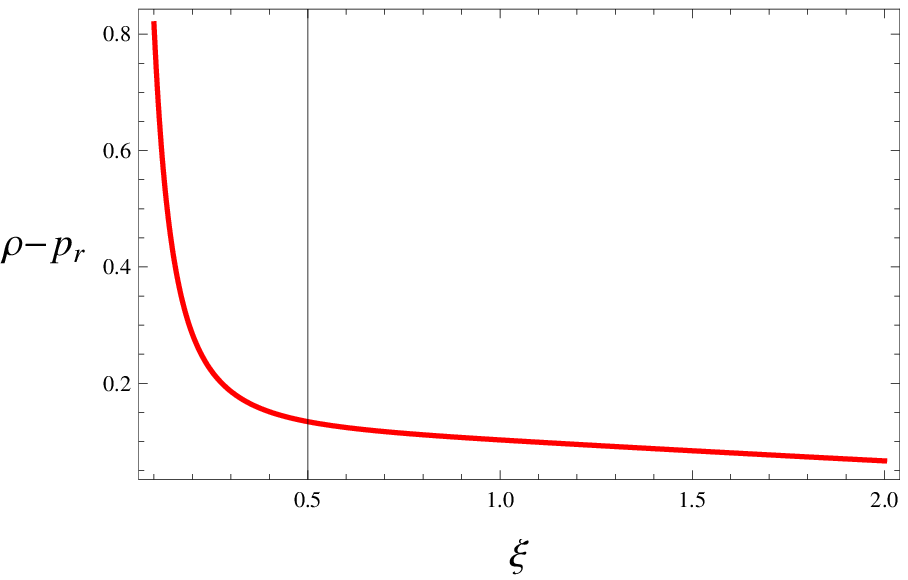,width=0.4\linewidth}
\epsfig{file=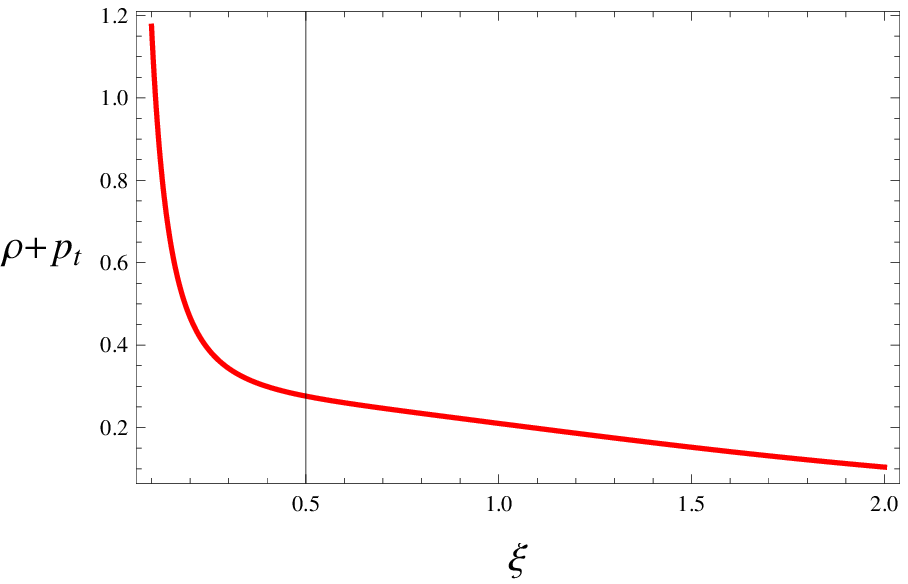,width=0.4\linewidth}\epsfig{file=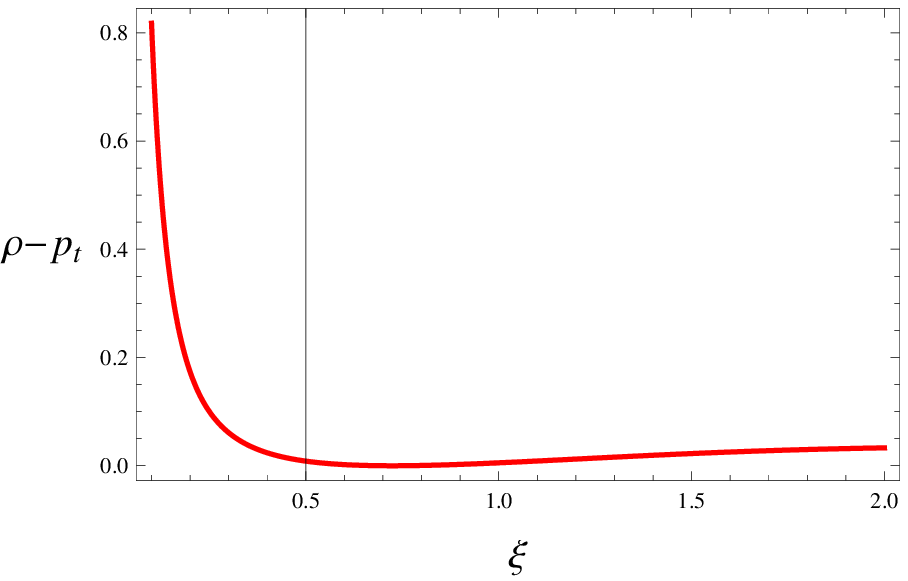,width=0.4\linewidth}
\epsfig{file=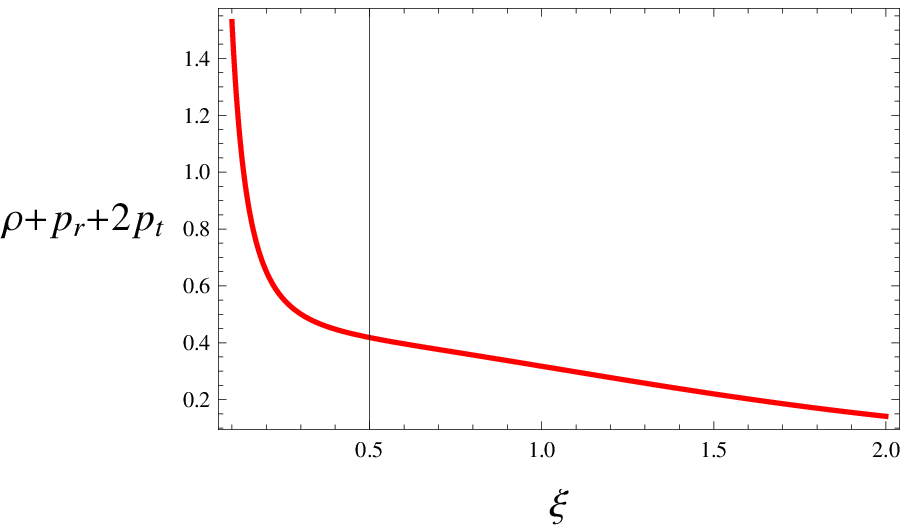,width=0.4\linewidth} \caption{Plots of
energy conditions for $\zeta=0.1$, $\rho_c=1$, $\mathcal{H}=0.9$ and
$\mathcal{N}=0.3$ corresponding to Model II.}
\end{figure}
\begin{figure}\center
\epsfig{file=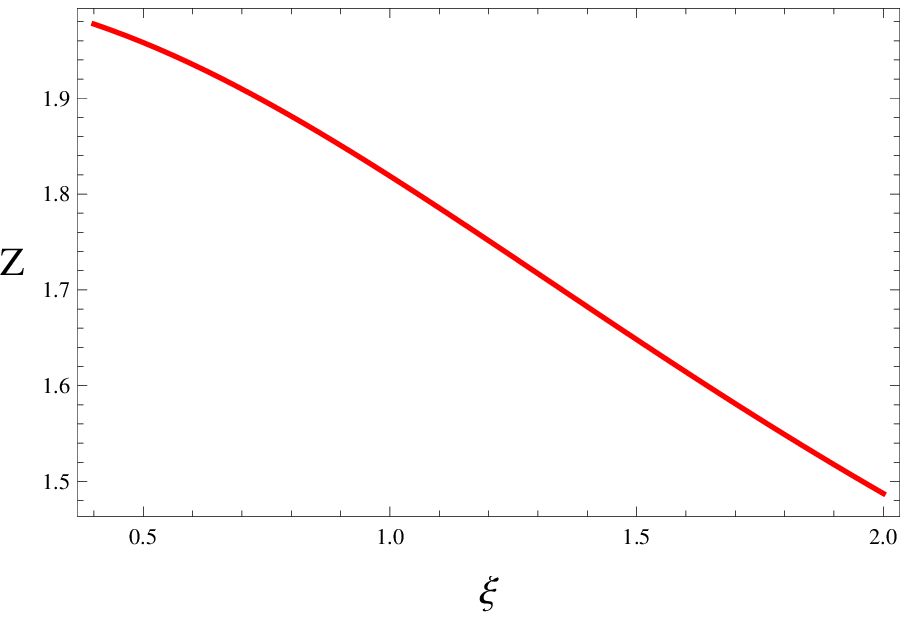,width=0.4\linewidth}\epsfig{file=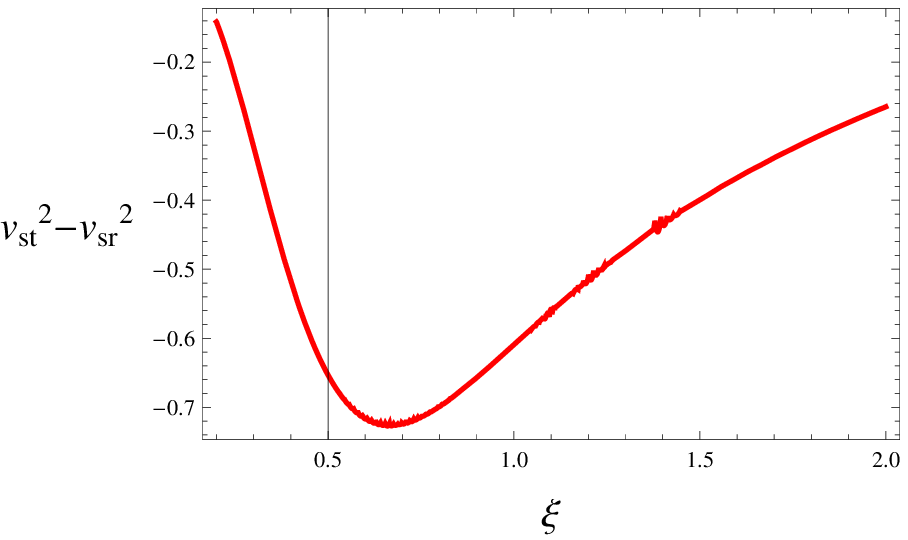,width=0.4\linewidth}
\caption{Plots of redshift and cracking condition for $\zeta=0.1$,
$\rho_c=1$, $\mathcal{H}=0.9$ and $\mathcal{N}=0.3$ corresponding to
Model II.}
\end{figure}

\subsection{Complexity-free Condition with a Non-local Equation of State}

A non-local equation of state was firstly proposed in \cite{ai} to
discuss a self-gravitating system. Its expression is given in the
following
\begin{align}\label{g62}
p_r=\rho-\frac{2}{r^3}\int_0^r\rho\bar{r}^2d\bar{r}+\frac{\mathbb{C}}{2\pi
r^3},
\end{align}
where $\mathbb{C}$ is a constant. The above equation can be written
after using the mass function \eqref{g12b} as
\begin{align}\label{g63}
p_r=\frac{1}{4\pi{r^2}}\bigg(m'-\frac{ss'}{r}\bigg)-\frac{1}{2\pi{r^3}}\bigg(m-\int_0^r\frac{ss'}{\bar{r}}d\bar{r}\bigg)
+\frac{\mathbb{C}}{2\pi r^3}.
\end{align}
Equation \eqref{g63} along with $\mathcal{Y}_{TF}=0$ are the
necessary tools to solve the field equations \eqref{g8}-\eqref{g10}.
They provide a set of two differential equations in terms of metric
functions $\beta_1$ and $\beta_2$ which are solved numerically to
understand the nature of the corresponding solution.
\begin{figure}\center
\epsfig{file=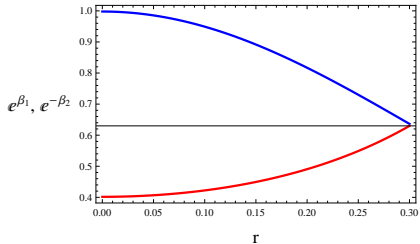,width=0.4\linewidth}
\caption{Plots of $e^{\beta_1}$ (red) and $e^{-\beta_2}$ (blue)
corresponding to Model III.}
\end{figure}

Figure \textbf{10} exhibits a singularity-free, positive and finite
behavior of $\beta_1$ and $\beta_2$, indicating physically
acceptable solution. Further, we have $e^{\beta_1(0)}=\mathcal{C}$
and $e^{-\beta_2(0)}=1$ at $r=0$, and they meet at a single point at
$r=0.3$. The compactness can now be calculated at the boundary,
since we have
\begin{figure}\center
\epsfig{file=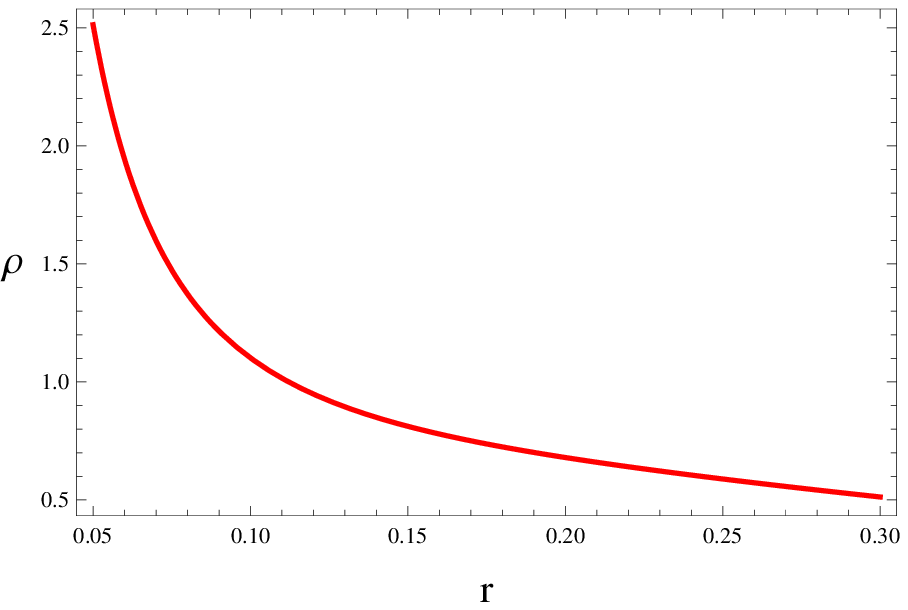,width=0.4\linewidth}\epsfig{file=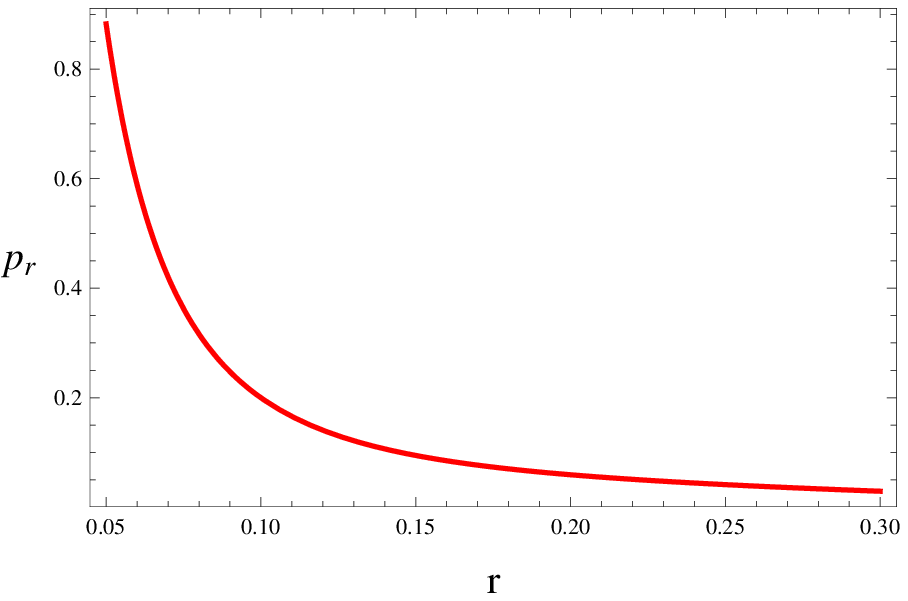,width=0.4\linewidth}
\epsfig{file=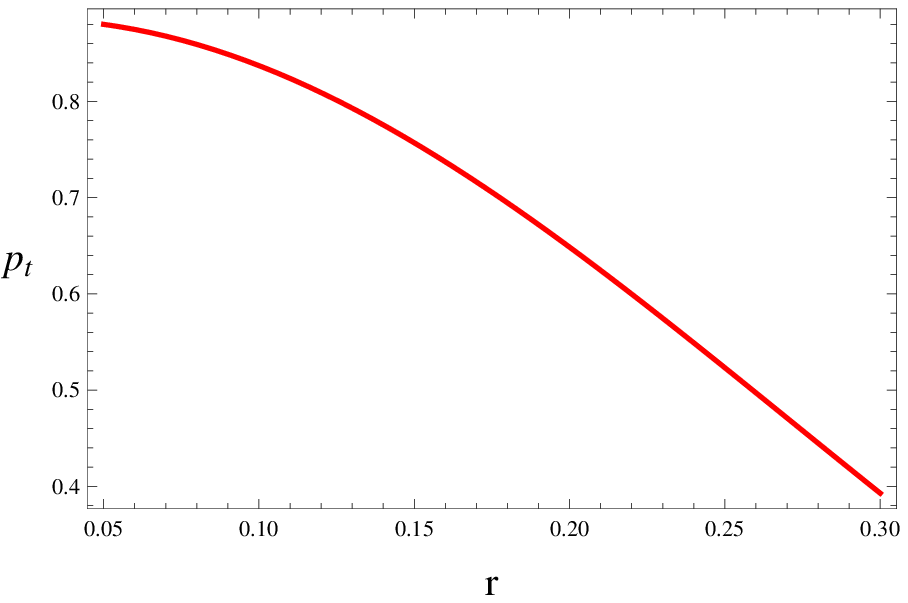,width=0.4\linewidth}\epsfig{file=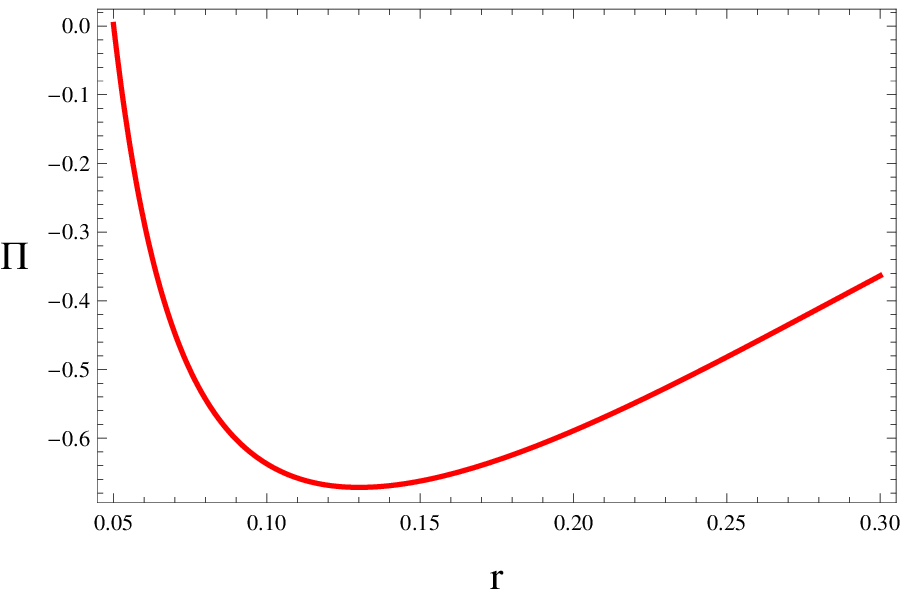,width=0.4\linewidth}
\caption{Plots of energy density, radial/tangential pressures and
anisotropy corresponding to Model III.}
\end{figure}
$$e^{\beta_1(0.3)}=e^{-\beta_2(0.3)}\approx0.63=1-\frac{2\mathcal{\bar{M}}}{\mathcal{R}}
+\frac{\mathcal{\bar{S}}^2}{\mathcal{R}^2},$$ which gives
\begin{align}\label{g64}
\frac{2\mathcal{\bar{M}}}{\mathcal{R}}
-\frac{\mathcal{\bar{S}}^2}{\mathcal{R}^2}\approx0.37 < \frac{8}{9}.
\end{align}
Figure \textbf{11} indicates acceptable nature of physical variables
as they are maximum and positive at the center and then decrease
outwards. Here, $p_t>p_r$ is observed outwards leading to the
negative anisotropy (lower right plot). The energy bounds are
demonstrated in Figure \textbf{12} which are satisfied, therefore,
the developed model is viable and contains ordinary matter. Figure
\textbf{13} (left) manifests the interior redshift that decrease
with $r$ and obtain the value at the boundary as $Z(0.3)\approx1.76
< Z_{\mathcal{R}}$. The cracking condition in this case takes
negative values only for small radius and then becomes positive
(right plot). Hence, our developed solution corresponding to the
constraints \eqref{g34} and \eqref{g62} is not stable dissimilar to
the uncharged case \cite{40}.
\begin{figure}\center
\epsfig{file=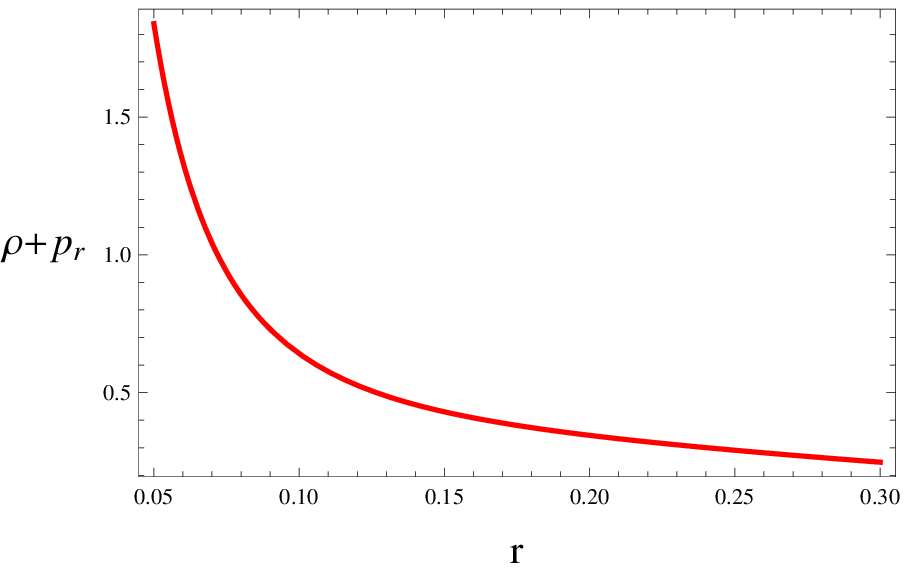,width=0.4\linewidth}\epsfig{file=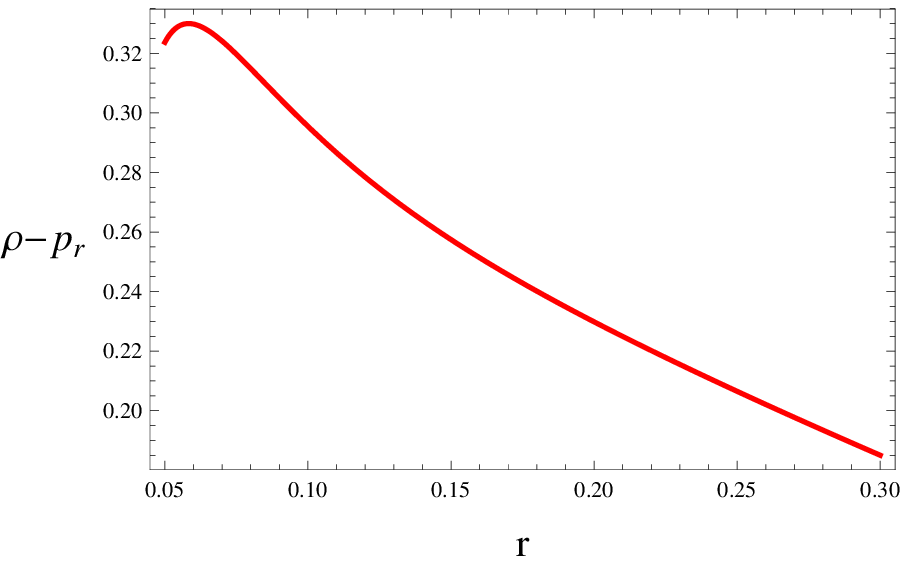,width=0.4\linewidth}
\epsfig{file=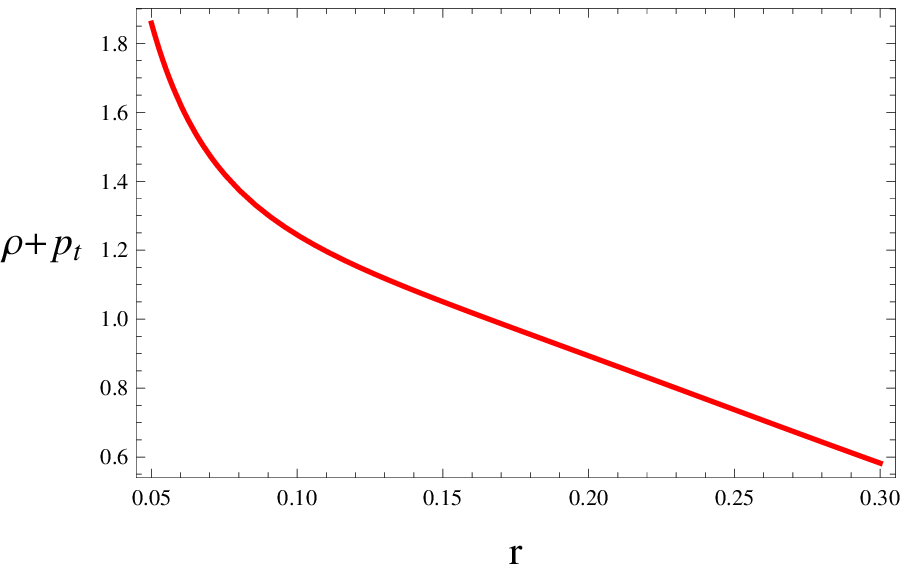,width=0.4\linewidth}\epsfig{file=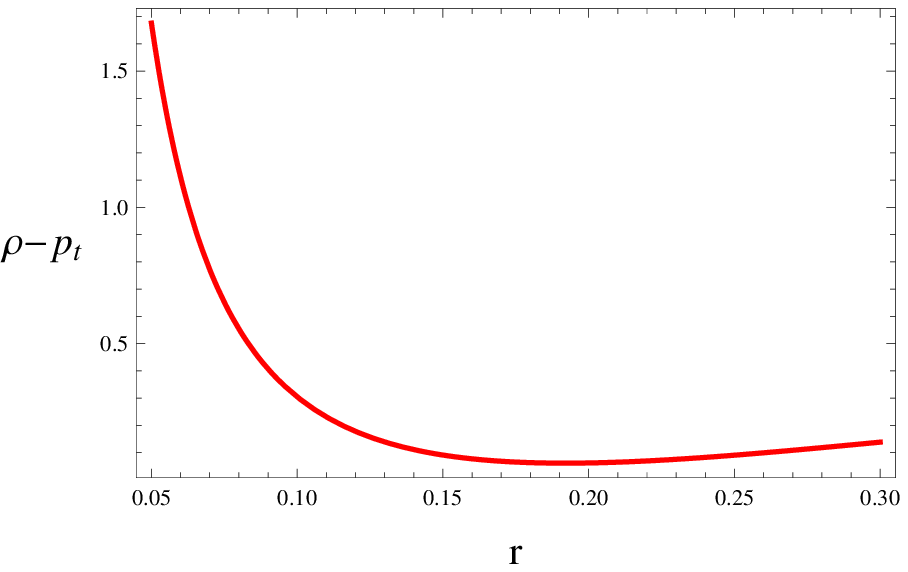,width=0.4\linewidth}
\epsfig{file=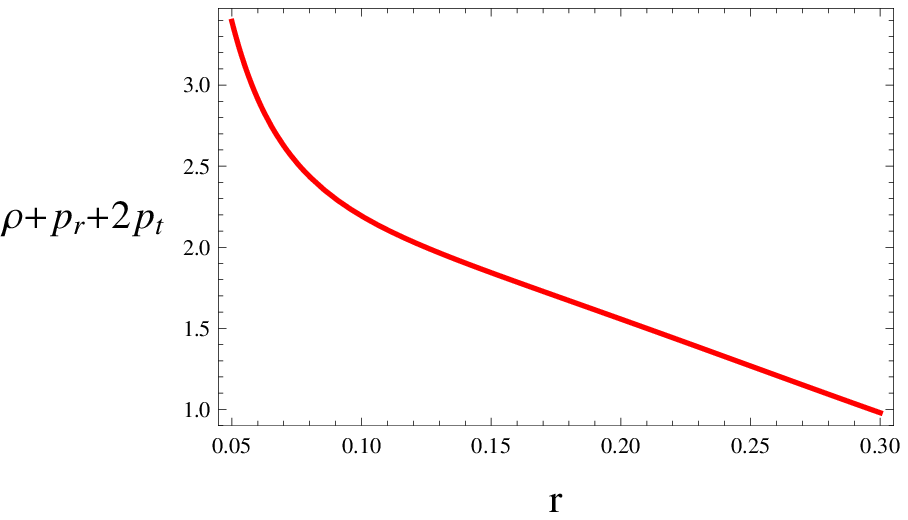,width=0.4\linewidth} \caption{Plots of
energy conditions corresponding to Model III.}
\end{figure}
\begin{figure}\center
\epsfig{file=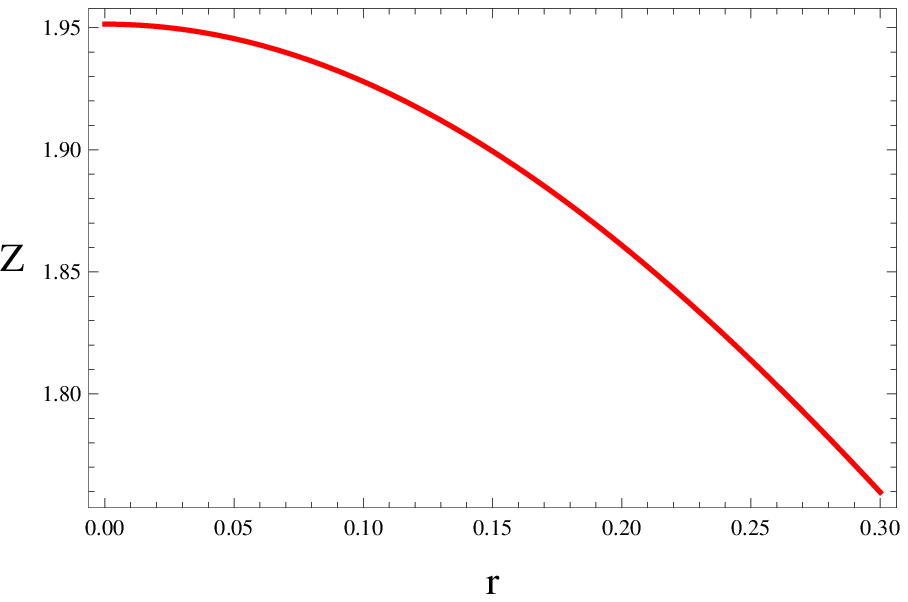,width=0.4\linewidth}\epsfig{file=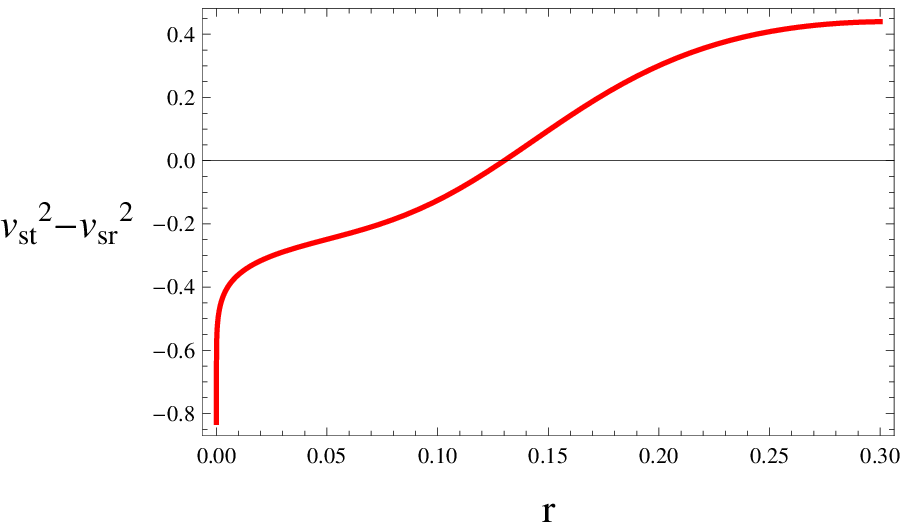,width=0.4\linewidth}
\caption{Plots of redshift and cracking condition corresponding to
Model III.}
\end{figure}

\section{Conclusions}

The purpose of this article is to formulate some different solutions
to the Einstein field equations in the presence of an
electromagnetic field. For this purpose, we have assumed a static
spherically symmetric interior geometry and determined the
Einstein-Maxwell field equations as well as the hydrostatic
equilibrium condition. We have also expressed mass of the sphere in
terms of energy density and charge. The Reissner-Nordstr\"{o}m
metric has then been taken as an exterior geometry to calculate
matching conditions at the spherical boundary. Further, we have
split the curvature tensor orthogonally and obtained four distinct
scalars associated with several physical parameters. We have
observed that a factor $\mathcal{Y}_{TF}$ involved density
inhomogeneity, anisotropy and charge, thus adopted it as the
complexity factor for the considered matter setup according to the
Herrera's suggested definition \cite{25}.

Since the field equations \eqref{g8}-\eqref{g10} contain five
unknowns, i.e., three matter variables and two metric potentials, we
have considered some constraints to make the system solvable. The
first of them has been taken as the complexity-free condition given
in Eq.\eqref{g34}. Moreover, we have chosen three constraints
($p_r=0$, a polytropic and a non-local equation of state,
respectively) as the second condition, leading to distinct
solutions. The solutions for $\beta_1$ and $\beta_2$ have been
calculated through numerically integrating the corresponding
equations in each case along with some initial conditions. We have
then presented some physical conditions whose fulfilment leads to
realistic stellar models. The matter sector (energy density and
pressure) corresponding to each solution has shown accepted behavior
(maximum at $r=0$ and decreasing outward). The compactness and
surface redshift have also been noticed to be less than their upper
limits. All the solutions have met viability criterion as the energy
conditions are satisfied. The cracking condition has been fulfilled
only by the solutions corresponding to $p_r=0$ and a polytropic
equation of state, however, it has taken positive values for the
case of third model leading to instability of that solution. It is
mentioned here that our resulting solutions corresponding to second
and third models are not consistent with \cite{40}.\\\\
\textbf{Data Availability:} No data was used for the research
described in this paper.

\end{document}